\documentclass{elsarticle}

\newtheorem{definition}{\textbf{Definition}}

\newtheorem{example}{\textbf{Example}}

\begin{document}

\title{\textbf{Behavior-aware Service Access Control Mechanism using Security Policy Monitoring for SOA Systems}}

\author[1]{Yunfei Meng}
\author[1]{Zhiqiu Huang}
\author[1]{Senzhang Wang}
\author[1]{Yu Zhou}
\author[1]{Guohua Shen}
\author[2]{Changbo Ke}

\address[1]{College of Computer Science and Technology, Nanjing University of Aeronautics and Astronautics, Nanjing 211106, China}
\address[2]{School of Computer Science and Technology, Nanjing University of Posts and Telecommunications, Nanjing 210023, China}

\begin{abstract}
Service-oriented architecture (SOA) system has been widely utilized at many present business areas. However, SOA system is loosely coupled with multiple services and lacks the relevant security protection mechanisms, thus it can easily be attacked by unauthorized access and information theft. The existed access control mechanism can only prevent unauthorized users from accessing the system, but they can not prevent those authorized users (insiders) from attacking the system. To address this problem, we propose a behavior-aware service access control mechanism using security policy monitoring for SOA system. In our mechanism, a monitor program can supervise consumer's behaviors in run time. By means of trustful behavior model (TBM), if finding the consumer's behavior is of misusing, the monitor will deny its request. If finding the consumer's behavior is of malicious, the monitor will early terminate the consumer's access authorizations in this session or add the consumer into the Blacklist, whereby the consumer will not access the system from then on. In order to evaluate the feasibility of proposed mechanism, we implement a prototype system. The final results illustrate that our mechanism can effectively monitor consumer's behaviors and make effective responses when malicious behaviors really occur in run time. Moreover, as increasing the rule's number in TBM continuously, our mechanism can still work well.
\end{abstract}


\maketitle

\section{Introduction}

\paragraph{}
In cloud computing era, massive free cloud services and low-cost computing resources are driving more and more enterprises to migrate their business processes into the cloud for the purpose of maximizing their profits. As an important enabling technique, service-oriented architecture (SOA) allows service consumers to build their flexible customized systems which integrate multiple piece of component services into one big service composition \cite{SOA1}. Since can implement the interconnection between heterogeneous systems with low cost, SOA system has been widely utilized at present different business areas, such as healthcare, education, government and e-commerce \cite{SOA2}. Especially with the rapid development of Internet of Things (IoT), SOA-based IoT has been proposed and attracted more attentions in recent years \cite{IOT1}\cite{IOT2}. However, SOA system is loosely coupled with multiple cloud services or some device services deployed in edge servers \cite{IOT1}, lacks the relevant security protection mechanisms, thus it can easily be attacked by unauthorized access and information theft. In SOA system, service requests may come from service consumers, intermediate services or both of them. Consumers and intermediate services may be distributed in different trust domains, thus service invocations across trust domains will incur new challenges for service access authorization management. For service consumers, the services they invoke probably need to access the other services to fulfill some additional requirements. In this case, the sensitive data carried in the request may be forwarded to the other distrusted services, which inevitably may lead to the potential information leakage for service consumers \cite{SOA3}.

\paragraph{}
The access control mechanism has been introduced to solve the problems of illegal and unauthorized access in SOA system. But traditional access control mechanisms, such as identity-based access control (IBAC) \cite{SOA4}, role-based access control (RBAC) \cite{SOA5}, are vulnerable when facing insider attacks, i.e., insiders misuse their privileges for malicious purposes \cite{insider1}. That is, traditional access control mechanism can only prevent unauthorized users from accessing the system, but due to lacking effective user behavior monitoring mechanism, they can not prevent those authorized users (insiders) from attacking the system, e.g., high-frequency access to services incurs the paralysis of entire system. To address these problems, runtime monitoring and assertion checking need to be introduced as means to enforce the security of SOA system \cite{SOA6}. Runtime execution monitoring (REM) is a class of methods of tracking the temporal behavior of an underlying application, REM methods range from simple print statement logging methods to run-time tracking of complete formal requirements for verification purposes \cite{SOA7}.

\paragraph{}
Based on these insights, we propose a behavior-aware service access control mechanism using security policy monitoring for SOA system. The primary goal our mechanism wants to reach has two perspectives. The first one is to implement fine-grained service access control by defining service releasing policy. Service releasing policy is defined by service provider, which strictly regulate which service in SOA system could be accessed by which service consumer for what target. Another perspective is to implement runtime monitoring by using trustful behavior model (TBM) and behavior-aware access control algorithm. By means of TBM, behavior-aware access control algorithm run in the monitor can supervise consumer's behaviors in run time. When consumer's behavior is judged to be malicious behavior, the monitor will adjust the consumer's access authorizations and prevent the consumer from continuing to use the system automatically.

\paragraph{}
In summary, this paper makes the following contributions: (1) We propose a system model by which we can formalize a SOA system. (2) Based on the system model, we propose a behavior-aware service access control mechanism using security policy monitoring, which consists of service releasing policy (SRM), trustful behavior model (TBM), behavior risk measurement models and behavior-aware access control algorithm. (3) We implement a proof-of-concept experimental system, whereby we can evaluate the effectiveness and performance of our proposed mechanism.

\paragraph{}
The remainder of the paper is structured as follows. Section 2 presents the system model. Section 3 is the main body of this paper, where we presents our proposed mechanism in detail. Section 4 implements the mechanism and evaluates its effectiveness and performance. Section 5 reviews some related works regarding to runtime monitoring for SOA system. Finally, Section 6 concludes this paper and presents some future directions.

\section{System Model}

\paragraph{}
In this paper, we only consider the SOA systems with single trust domain, i.e., if a service provider is the owner of SOA system, the service provider can define the security policy for entire SOA system. Following, we first establish the accurate formal definitions for modeling service provider, service consumer and SOA system respectively. Then, based on these definitions, we utilize an example to illustrate how to establish the SOA system model in detail.

\begin{definition}(Service Provider):
\emph{
Service provider is the owner of SOA system. It is defined as follows. $SP$=$\langle$ $P_{i}$, $PK_{i}$ $\rangle$, where: $P_{i}$ is the identity of service provider in SOA system. $PK_{i}$ is the private key of service provider in SOA system.
}
\end{definition}

\begin{definition}(Service Consumer):
\emph{
Service consumer is the user of SOA system. It is defined as follows. $SC$=$\langle$ $C_{i}$, $CK_{i}$ $\rangle$, where: $C_{i}$ is the identity of service consumer in SOA system. $CK_{i}$ is the private key of service consumer in SOA system.
}
\end{definition}

\begin{definition}(SOA System):
\emph{
SOA system is a service composition with single trust domain. It is defined as follows. $SOA$=$\langle$ $\mathcal{S}$, $\mathcal{T}$, $\mathcal{I}_{0}$, $\mathcal{UR}$, $\mathcal{L}$ $\rangle$, where: $\mathcal{S}$=$\mathcal{SS}$$\cup$$\mathcal{DS}$ is a finite set of services. $\mathcal{SS}$ is a finite set of system services, $\mathcal{DS}$ is a finite set of sensitive services. System services can be accessed by any consumers, while sensitive services are those services which have massive sensitive data or privacies, thus they are protected by system and can only be accessed by those authorized users who are defined by service provider directly. $\mathcal{T}\subseteq\mathcal{S}\times\mathcal{S}$ is a finite set of directed transitions. If $\exists s_{i},\ s_{j}\in\mathcal{S}$, then the pair $t_{k}$=$\{s_{i}$, $s_{j}\}$ is a transition from service $s_{i}$ to service $s_{j}$. $\mathcal{I}_{0}\in\mathcal{S}$ is the initial service of SOA system. $\mathcal{UR}$ is a finite set of uniform resource identifier (URI) of service interface. $\mathcal{L}$ is a function $\mathcal{L}:$ $\mathcal{S}\rightarrow\mathcal{UR}$, and $\forall s_{i}, s_{j}\in\mathcal{S}$, $\mathcal{L}(s_{i})\neq\mathcal{L}(s_{j})$. And if $u_{i}\in\mathcal{UR}$, $s_{j}\in\mathcal{S}$, then $\mathcal{L}(s_{j})$=$u_{i}$ represents the URI of $s_{j}$'s interface is at $u_{i}$.
}
\end{definition}

\begin{example}
The electronic medical records sharing system (EMRSS) is a typical SOA system used in healthcare. By means of EMRSS, medical facility can manage their patients' electronic medical records (EMR) and share these records with medical experts for disease treatments. We assume that Bob is the service provider of an EMRSS. The EMRSS can provide two categories of sensitive services, i.e., electrocardiogram (ECG) services and X-Ray image services. ECG services can be further divided into ECG online exploring service and ECG updating service. X-Ray image service can only provide image online exploring service. If we let $S_{1}$ is X-Ray image online exploring service, $S_{3}$ is ECG online exploring service, $S_{4}$ is ECG updating service, then the EMRSS can be modelled as a SOA system model shown in Figure 1. Here $\mathcal{SS}$=$\{$$S_{0}$, $S_{2}$$\}$, $\mathcal{DS}$=$\{$$S_{1}$, $S_{3}$, $S_{4}$$\}$, $\mathcal{T}$=$\{$$t_1$, $t_2$, $t_3$, $t_4$$\}$, $\mathcal{I}_{0}$=$\{$ $S_0$ $\}$, $\mathcal{UR}$=$\{$ /SBA/0.jsp, /SBA/X0.jsp, /SBA/1.jsp, /SBA/X1.jsp, /SBA/X2.jsp $\}$, $\mathcal{L}$=$\{$$\mathcal{L}(S_{0})$= /SBA/0.jsp, $\mathcal{L}(S_{1})$= /SBA/X0.jsp, $\mathcal{L}(S_{2})$= /SBA/1.jsp, $\mathcal{L}(S_{3})$= /SBA/X1.jsp, $\mathcal{L}(S_{4})$= /SBA/X2.jsp$\}$.
\end{example}

\section{Behavior-aware Service Access Control mechanism}

\paragraph{}
Based on the SOA system model mentioned above, we propose a behavior-aware service access control mechanism using security policy monitoring. In some sense, this mechanism can be thought as a combination of fine-grained service access control and runtime monitoring. In the following of this section, we first present the entire mechanism. Then we introduce two important formal model, service releasing model and trustful behavior model. Next, we present a group risk quantitative measurement models for evaluating user's behavior risks in SOA system. Finally, we give the detailed algorithmic description of behavior-aware access control algorithm in our mechanism.

\subsection{Mechanism Description}

\paragraph{}
The proposed mechanism can be described as a flow chart shown in Figure 2. The primary goal our mechanism wants to reach has two perspectives, The first one is to implement fine-grained service access control by defining service releasing policy. Another perspective is to implement runtime monitoring by using trustful behavior model and behavior-aware access control algorithm.

\paragraph{}
Thus, we design an architecture based on check points and a monitor for our mechanism. Check point is a section of script code embedded in each service interface of SOA system, which is responsible for collecting service consumer (SC)'s behavior elements, such as identity, location, IP address and etc. Monitor is designed as an independent supervising program which is responsible for monitoring SC's behaviors in run time.

\paragraph{}
Firstly, service provider (SP) defines a service releasing policy (SRM), which is stored in the monitor and strictly regulates which sensitive service in system could be released to which SC for what target. Here releasing a sensitive service refers to the sensitive service could be accessed or operated by SC. Then SRM will be converted into a group corresponding trustful behavior models (TBM) for different SCs. Each TBM is a set of trustful behavior rules which are taken as the criterion of behavior-aware access control algorithm. After that, once a service invoked by SC, the monitor will be activated automatically. The behavior elements of SC collected by check point will be sent to the monitor for analysing. If SC's access request doesn't meet TBM, monitor will deny SC's request because the current behavior is unpermitted. If finding SC's behaviors have malicious intentions, i.e., the new behavior risk calculated by risk measurement models has exceeded its risk threshold, monitor will early terminate the SC's access authorization in this session for guaranteeing the security of system. In this case, SC will not access any services in system, but also can access the system in next new session.

\subsection{Service Releasing Policy}

\paragraph{}
Leveraging P-Spec policy specification language we proposed in reference \cite{meng}, we can formalize SP's service releasing policy as formal P-Spec policy model. By means of SRM, SP can precisely define which sensitive service could be released to which SC for what target. The relative definitions are as follows.

\begin{definition}(service Releasing Rule):
\emph{
Given system $SOA$, $\mathcal{DS}\subset$$SOA$. Each service releasing rule $RR_{i}$ is defined as follows: $\mathcal{I}(\mathcal{N})$= $\{Authentication$, $Authorization\}$. $\mathcal{I}(\mathcal{K})$=$\{SC$, $CK$, $Target$, $Service\}$, where $SC$ represents a service consumer, $CK$ represents the SC's private key in system, $Target$ represents a specific purpose, $Service$ represents a sensitive service which could be released to the $SC$ for $Target$. $\mathcal{I}(\mathcal{V})$=$\{C_{i}$, $CK_{i}$, $T_{i}$, $X_{i}\}$, where $C_{i}$ is the corresponding variable of keyword $DU$, $CK_{i}$ is the corresponding variable of keyword $CK$, $T_{i}$ is the corresponding variable of keyword $Target$, $X_{i}$ is the corresponding variable of keyword $Serivce$, and $\forall X_{i}\in\mathcal{DS}$. $\mathcal{I}(\mathcal{R})$ is defined as follows:
\begin{equation}
\begin{array}{l}
RR_{i}::=Authentication\rightarrow Authorization\\
Authentication::=(SC=C_{i})\&\&(CK=CK_{i})\\
\ \ \ \ \ \ \ \ \ \ \ \ \ \ \&\&(Target=T_{j})\\
Authorization::=(Service=X_{k})\\
\ \ \ \ \ \ \ \ \ \ \ \ \ \ |\ Authorization\ ||\ Authorizaiton\\
\end{array}
\end{equation}
Where The symbol = represents semantical \emph{equal}, the symbol $||$ represents logical \emph{or}, and the symbol $\rightarrow$ represents logical \emph{imply to}. Semantically, each $RR_{i}$ regulates if service consumer $C_{i}$ whose private key $CK_{i}$ and target $T_{j}$ can pass the authentication, then $C_{i}$ will be granted the relevant authorizations to access or operate these sensitive services $\{X_{i}\}$ accordingly.
}
\end{definition}

\begin{definition}(Service Releasing Model):
\emph{
Service Releasing model is defined as a set of $RR_{i}$ rules, we denote it as: $SRM=\{RR_{0},RR_{1},...,RR_{n}\}$.
}
\end{definition}

\begin{example}
Following the background of example 1, we assume that Dr. Mike is a cardiologist, he requires patient's electrocardiogram (ECG) and chest X-ray images for treating cardiopathy. Dr. Mary is an internist, she requires patient's chest X-ray images for treating influenza. Thus, both Mike and Mary are of service consumers. As service provider, Bob has authorized Dr. Mike and Dr. Mary to access the EMRSS. Due to all ECG and X-Ray images being of sensitive privacies for patients, thus Mike and Mary can't download or update any data except exploring them online. Hence, SRM can be defined as follows. Firstly, let $C_{0}$ = Mike, $CK_{0}$ = Mike's private key in system, $C_{1}$ = Mary, $CK_{1}$ = Mary's private key in system, $T_{0}$ = cardiopathy, $T_{1}$ = influenza, $X_{0}$ = X-Ray exploring service, $X_{1}$ = ECG exploring service. Then, SRM is the set $\{$$RR_{0}$, $RR_{1}$$\}$ and shown in Table 1.
\begin{table}[htbp]
\caption{The established SRM.}
\begin{center}
\begin{tabular}{l|l}
\hline
\scriptsize{$\ RR_{0}\ $} & \scriptsize{$(SC=C_{0})\&\&(CK=CK_{0})\&\&(Target=T_{0})\rightarrow(Service=X_{0})||(Service=X_{1})$}\\
\hline
\scriptsize{$\ RR_{1}\ $} & \scriptsize{$(SC=C_{1})\&\&(CK=CK_{1})\&\&(Target=T_{1})\rightarrow(Service=X_{0})$}\\
\hline
\end{tabular}
\end{center}
\end{table}
\end{example}

\subsection{Trustful Behavior Model}

\paragraph{}
In some sense, our SRM is similar with the existed role-based access control (RBAC) or attribute-based access control (ABAC). RBAC or ABAC can also authenticate a SC's legitimacy, and grant the SC with relevant authorizations. But they can't monitor an authenticated SC's behaviors in run time. Hence, we introduce a trustful behavior model (TBM), which is converted from SRM and can assist the monitor to supervise an authenticated users' behaviors in run time. The relative definitions are as follows.

\begin{definition}(Trusted Behavior Rule):
\emph{
Each trustful behavior rule $RB_{i}$ is defined as a 3-tuple: $\langle Id$, $Src$, $Dst\rangle$, where $Id$ is the identification of a SC, $Src$ is the URI of current service interface, $Dst$ is the URI of target service interface. Semantically, each $RB_{i}$ rule represents if a SC whose identity equals $Id$ wants to invoke the target service $Dst$ from the current service $Src$, then the behavior is believed trustful for SP, otherwise it is distrustful.
}
\end{definition}

\begin{definition}(Model Conversion):
\emph{
Given system model $SOA$ and $SRM$, $\exists RR_{j}\in SRM$ whose $SC$=$C_{i}$. If $\forall X_{k}\in RR_{j}$, there still exists a route $rt_{k}$=$\{$ $t_{0}$, $t_{1}$,...,$t_{m}$ $\}\subset\mathcal{T}$ in $SOA$, whose initial service is $\mathcal{I}_{0}\in$ $SOA$ and the end service is $X_{k}$, then set $\{$$C_{i}$, $t_{i}$$\}$ can be converted into a group of corresponding trustful behavior rules $\{RB_{X_{k}}^{t_{i}}\}$, and the conversion procedure can be defined as a function.
\begin{equation}
\{C_{i},\ t_{i}\}\rightarrow\{RB_{X_{k}}^{t_{i}}\}=\left\{
\begin{array}{l}
\langle C_{i},\ \mathcal{L}(S_{p}),\ \mathcal{L}(S_{q}) \rangle \\
\langle C_{i},\ \mathcal{L}(S_{p}),\ \mathcal{L}(S_{p}) \rangle \\
\langle C_{i},\ \mathcal{L}(S_{q}),\ \mathcal{L}(S_{q}) \rangle \\
\end{array}
\right.
\end{equation}
Where $t_{i}$=$\{$$S_{p}$, $S_{q}$$\}\in rt_{k}$, $\mathcal{L}(S_{p})\in\mathcal{L}$ is the URI of $S_{p}$'s service interface, $\mathcal{L}(S_{q})\in\mathcal{L}$ is the URI of $S_{q}$'s service interface. Semantically, equation (2) represents the set $\{C_{i},\ t_{n}\}$ can be converted into three trustful behavior rules respectively, the first rule represents the behavior of invoking service $S_{q}$ from service $S_{p}$, the second rule represents the behavior of refreshing $S_{p}$'s client, and the third one represents the behavior of refreshing $S_{q}$'s client.
}
\end{definition}

\begin{definition}(Trusted Behavior Model):
\emph{
Given system model $SOA$ and $SRM$, $\exists RR_{j}\in SRM$ whose $SC$=$C_{i}$ has releasing service set $\mathcal{X}$=$\{$ $X_{0}$, $X_{1}$,...,$X_{n}$ $\}$. If $\forall X_{k}\in\mathcal{X}$, there still exists a route $rt_{k}$=$\{$ $t_{0}$, $t_{1}$,...,$t_{m}$ $\}\subset\mathcal{T}$ in $SOA$, whose initial service is $\mathcal{I}_{0}\in$ $SOA$ and the end service is $X_{k}$, then trusted behavior model of $C_{i}$ denoted as $TBM^{C_{i}}$ can be converted from set $\mathcal{X}$ and set $\mathcal{RT}$=$\{$ $rt_{0}$, $rt_{1}$,...,$rt_{n}$ $\}$, and the conversion procedure can be defined as a function.
\begin{equation}
TBM^{C_{i}}=\displaystyle\bigcup_{k=0}^{n}(\bigcup_{i=0}^{m}\ \{RB_{X_{k}}^{t_{i}}\})\\
\end{equation}
Where $t_{i}\in rt_{k}$, $rt_{k}\in\mathcal{RT}$, $X_{k}\in\mathcal{X}$, $|\mathcal{X}|$=$|\mathcal{RT}|$. Here $\{RB_{X_{k}}^{t_{i}}\}$ is the set of all trusted behavior rules converted from $SOA$ and $SRM$ using equation (2).
}
\end{definition}

\paragraph{}
Based on SRM in Table 1 and SOA system model in Figure 1, we can create the trustful behavior model for Dr.Mike and Dr. Mary respectively. Firstly, according to SRM, releasing service set of Mike is $\{$$X_{0}$, $X_{1}$$\}$, and releasing service of Mary is $\{$$X_{0}$$\}$. Then in SOA, the route from $\mathcal{I}_{0}$ to $X_{0}$ is transition set $\{$$t_{1}$$\}$, the route from $\mathcal{I}_{0}$ to $X_{1}$ is transition set $\{$$t_{2}$, $t_{3}$$\}$, and the entire derivation process can be depicted in Figure 3. Next using equation (2) and (3), TBM of Mike and Mary can be established and shown in Table 2 and Table 3 respectively.

\begin{table}[htbp]
\caption{Trusted behavior model of \emph{Mike}.}
\begin{center}
\begin{tabular}{l|l}
\hline
\scriptsize{$\ \ \ RB_{0}\ \ \ $} & \scriptsize{$\ \ \ \langle\ C_{0},\ \ /SBA/0.jsp,\ \ /SBA/X0.jsp\ \rangle\ \ \ $}\\
\hline
\scriptsize{$\ \ \ RB_{1}\ \ \ $} & \scriptsize{$\ \ \ \langle\ C_{0},\ \ /SBA/0.jsp,\ \ /SBA/0.jsp\ \rangle\ \ \ $}\\
\hline
\scriptsize{$\ \ \ RB_{2}\ \ \ $} & \scriptsize{$\ \ \ \langle\ C_{0},\ \ /SBA/X0.jsp,\ \ /SBA/X0.jsp\ \rangle\ \ \ $}\\
\hline
\scriptsize{$\ \ \ RB_{3}\ \ \ $} & \scriptsize{$\ \ \ \langle\ C_{0},\ \ /SBA/0.jsp,\ \ /SBA/1.jsp\ \rangle\ \ \ $}\\
\hline
\scriptsize{$\ \ \ RB_{4}\ \ \ $} & \scriptsize{$\ \ \ \langle\ C_{0},\ \ /SBA/1.jsp,\ \ /SBA/1.jsp\ \rangle\ \ \ $}\\
\hline
\scriptsize{$\ \ \ RB_{5}\ \ \ $} & \scriptsize{$\ \ \ \langle\ C_{0},\ \ /SBA/1.jsp,\ \ /SBA/X1.jsp\ \rangle\ \ \ $}\\
\hline
\scriptsize{$\ \ \ RB_{6}\ \ \ $} & \scriptsize{$\ \ \ \langle\ C_{0},\ \ /SBA/X1.jsp,\ \ /SBA/X1.jsp\ \rangle\ \ \ $}\\
\hline
\end{tabular}
\end{center}
\end{table}

\begin{table}[htbp]
\caption{Trusted behavior model of \emph{Mary}.}
\begin{center}
\begin{tabular}{l|l}
\hline
\scriptsize{$\ \ \ RB_{0}\ \ \ $} & \scriptsize{$\ \ \ \langle\ C_{1},\ \ /SBA/0.jsp,\ \ /SBA/X0.jsp\ \rangle\ \ \ $}\\
\hline
\scriptsize{$\ \ \ RB_{1}\ \ \ $} & \scriptsize{$\ \ \ \langle\ C_{1},\ \ /SBA/0.jsp,\ \ /SBA/0.jsp\ \rangle\ \ \ $}\\
\hline
\scriptsize{$\ \ \ RB_{2}\ \ \ $} & \scriptsize{$\ \ \ \langle\ C_{1},\ \ /SBA/X0.jsp,\ \ /SBA/X0.jsp\ \rangle\ \ \ $}\\
\hline
\end{tabular}
\end{center}
\end{table}

\subsection{Risk Measurement Model}

\paragraph{}
In order to find and capture user's malicious behaviors during run time, we need to evaluate the user's behavior using quantitative risk measurement models. If a SC's behavior risk has exceeded its threshold, then we think the SC's behavior has malicious intentions. In this case, system will early terminate the SC's access authorization in this session. The relative risk measurement models are defined as follows.

\begin{definition}(Behavior Element):
\emph{
Behavior element of SC is an environment variable collected by check point embedded in service interface. The set of behavior elements is defined as $\mathcal{E}=\{$ $e_{1}$, $e_{2}$,...,$e_{n}\}$.
}
\end{definition}

\begin{definition}(Behavior Risk):
\emph{
Given a behavior element $e_{i}\in\mathcal{E}$, the risk of behavior $e_{i}$ is defined as a function.
\begin{equation}
R(e_{i},\ t_{i+1})=\mathcal{F}(e_{i},\ t_{i})\\
\end{equation}
The equation (4) represents the new behavior risk at time $t_{i+1}$ is determined by the existed risk at time $t_{i}$.
}
\end{definition}

\begin{definition}(Unauthorized Access Risk):
\emph{
Given $TBM^{C_{i}}$ and behavior element set $\mathcal{A}$=$\{$ $e_{id}$, $e_{src}$, $e_{dst}$ $\}$, $\mathcal{A}\subset\mathcal{E}$, where $e_{id}$ is SC's identify, $e_{src}$ is the URI of current service and $e_{dst}$ is the URI of target service. If $\exists RB_{j}\in TBM^{C_{i}}$, and $RB_{j}$=$\{$ $Id$, $Src$, $Dst$ $\}$, makes $e_{id}$=$Id$, $e_{src}$=$Src$, $e_{dst}$=$Dst$, then SC's unauthorized access risk (UAF) is defined as follows.
\begin{equation}
UAF=R(\mathcal{A},\ t_{i}) \\
\end{equation}
Otherwise, it is defined as follows.
\begin{equation}
UAF=R(\mathcal{A},\ t_{i})+1 \\
\end{equation}
Where $R(\mathcal{A},\ t_{i})$ is the existed unauthorized access risk.
}
\end{definition}

\begin{definition}(Access Retention Risk):
\emph{
If $e_{t}\in\mathcal{E}$ is the environment time, then access retention risk (ARR) is defined as follows.
\begin{equation}
ARR=R(e_{t},\ t_{i})+(t_{i+1}-t_{i}) \\
\end{equation}
Where $R(e_{t},\ t_{i})$ is the existed access retention risk.
}
\end{definition}

\begin{definition}(Access Frequency Risk):
\emph{
If $e_{f}\in\mathcal{E}$ is the times of SC accessing services, then access frequency risk (AFR) is defined as follows.
\begin{equation}
AFR=\frac{e_{f}(t_{i+1})-e_{f}(t_{i})}{t_{i+1}-t_{i}} \\
\end{equation}
where $e_{f}(t_{i+1})$ is the times of SC accessing services at the time $t_{i+1}$, $e_{f}(t_{i})$ is the times of SC accessing services at the time $t_{i}$.
}
\end{definition}

\begin{definition}(Risk Evaluation):
\emph{
Given a behavior element $e_{i}\in\mathcal{E}$, $R(e_{i},\ t_{i})$ is the behavior risk at the time $t_{i}$, then the risk evaluation towards $R(e_{i},\ t_{i})$ is defined as follows.
\begin{equation}
E(R(e_{i},\ t_{i}))=1-\frac{\theta(e_{i})}{R(e_{i},\ t_{i})}\\
\end{equation}
Where $\theta(e_{i})$ is the risk threshold of behavior $e_{i}$.
}
\end{definition}

\begin{definition}(Risk Evidence):
\emph{
Given a behavior element $e_{i}\in\mathcal{E}$, $R(e_{i},\ t_{i})$ is the behavior risk at the time $t_{i}$, the risk evaluation towards $R(e_{i},\ t_{i})$ is $E(R(e_{i},\ t_{i}))$, then the risk evidence towards $R(e_{i},\ t_{i})$ is defined as follows.
\begin{equation}
PF(R(e_{i},\ t_{i}))=\left\{
\begin{array}{l}
1,\ \ \ E(R(e_{i},\ t_{i}))>0  \\
0,\ \ \ E(R(e_{i},\ t_{i}))\leq0  \\
\end{array}
\right.
\end{equation}
Semantically, if $E(R(e_{i},\ t_{i}))>0$, it represents the accumulated risk $R(e_{i},\ t_{i})$ has exceeded the risk threshold at the time $t_{i}$, thus the risk evidence is set to 1 (true). If $E(R(e_{i},\ t_{i}))\leq 0$, then risk $R(e_{i},\ t_{i})$ is still acceptable at the time $t_{i}$, thus we set the evidence to 0 (false).
}
\end{definition}

\subsection{Behavior-aware Access Control Algorithm}

\paragraph{}
The behavior-aware access control algorithm can be described as a flow chart shown in Figure 4. According to the primary goal of our mechanism, the algorithm will implement two basic functions, the first one is to supervise SC's behaviors using $TBM^{C_{i}}$, another one is to evaluate SC's behavior risks in run time and makes the relevant decisions.

\paragraph{}
Thus, once the monitor session being activated by user $C_{i}$, the algorithm will first check the $C_{i}$'s behavior elements $\{$$e_{id}$, $e_{src}$, $e_{dst}$$\}$ using $TBM^{C_{i}}$. If $C_{i}$'s behavior elements don't meet $TBM^{C_{i}}$, algorithm will deny the request directly, otherwise it goes on to calculate $C_{i}$'s new behavior risks using the associated risk measurement models. Here we only consider two behavior risks, i.e., unauthorized access risk (UAR) and access frequency risk (AFR). If $PF(UAR)$=1 and $PF(AFR)$=1 at the same time, then $C_{i}$ is definitely a malicious user because s/he is trying to access those unauthorised services with abnormal access frequency. In this case, algorithm will add $C_{i}$ into the Blacklist and prohibit $C_{i}$ from then on. If finding $PF(UAR)=0$ and $PF(R(AFR)=0$ at the same time, that represents $C_{i}$'s behaviors have no any risks for SP, then algorithm will return the target service's interfaces to $C_{i}$ and update the existed risks with new calculated values. Otherwise, $C_{i}$'s behaviors still have risks, algorithm will deny $C_{i}$'s request in this session, but $C_{i}$ can also access the system in next new session.

\section{Experiment and Evaluation}

\paragraph{}
In this section, we implement a prototype to validate the feasibility of proposed mechanism. Following, we first introduce our experimental system. Then we further evaluate the effectiveness and performance of mechanism under different test scenarios.

\subsection{Experimental System}

\paragraph{}
In order to validate the feasibility of proposed mechanism, we implement a prototype system using Eclipse Java EE Neon3 and Apache Tomcat 7.0 container. All experiments are conducted on one PC using Intel Core i5 2.5GHz processors, running Windows 7 platform. The SOA system of designed prototype system is shown in Figure 5, which has 3 system services $\{$ $S_{0}$, $S_{1}$, $S_{2}$ $\}$ and 6 sensitive services $\{$ $S_{3}$, $S_{4}$, $S_{5}$, $S_{6}$, $S_{7}$, $S_{8}$ $\}$ respectively. Here we leverage java server page (JSP) to simulate the real service interface in system. In each JSP, we have embedded a section of check point script code which is responsible for collecting user's behavior elements in run time. We simulate the monitor using a Java Session Bean, and have implemented the behavior-aware access control algorithm mentioned in section 4.5 with Java.

\paragraph{}
For convenient of establishing different test scenarios rapidly, we also developed a simple policy management toolkit in our prototype system, whose screenshot is shown in Figure 6. By means of the toolkit, service providers can rapidly design different service releasing policy (SRM) and trustful behavior model (TBM) for different service consumers. Here there are totally 6 privacy services (Service 1 $\sim$ Service 6) which could be chosen for releasing to SC. \emph{Create} refers to creating a completely new TBM file, \emph{Append} refers to inserting some new trust behavior rules into the existed TBM file.

\subsection{Evaluation}

\paragraph{}
We first evaluate user behavior supervision, i.e., once the monitor session being activated, only access requests having satisfied with TBM could be responded, while others will be denied. Next we want to validate whether our mechanism can make effective responses when unauthorized access risk or access frequency risk really occur. Finally, we want to evaluate the performance of our mechanism.

\subsubsection{User Behavior Supervision}

\paragraph{}
This evaluation is to validate whether only access requests satisfied with TBM can be responded, while others can not. Specifically, we first design two test scenarios using the policy management toolkit shown in Figure 6. The first scenario ($SRM_{1}$) is defined as only Service 1 and Service 5 could be released to data users, and the second one ($SRM_{2}$) is defined as only Service 1, Service 3 and Service 4 could be released to data users.

\begin{figure}[htbp]
\centering
\scalebox{0.33}{\includegraphics{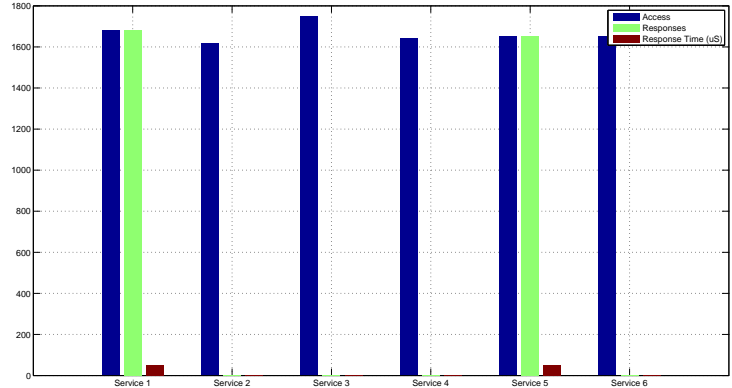}}
\caption{User behavior supervision evaluation under $SRM_{1}$.}
\end{figure}

\paragraph{}
Based on the SOA system model shown in Figure 5, $SRM_{1}$ and $SRM_{2}$ can be converted into their corresponding TBM models by policy management toolkit automatically. Next we randomly create 10000 access requests whose range covers from Service 1 to Service 6. We further design that if a service receives an access request, it will return a response to system, then we can record all responses time created by the services having been accessed successfully. After experiments finished, we can get the final result. Figure 7 shows the execution result under $SRM_{1}$ and Figure 8 shows the execution result under $SRM_{2}$. We can observe that, only access requests sent for Service 1 and Service 5 could be responded under $SRM_{1}$, while other requests have been denied by monitor. Taking Service 1 as an example in Figure 7, the access times is totally 1682, and all accesses for Service 1 have been responded and the average response time is 51 uS. Under the scenario of $SRM_{2}$, we find only access requests for Service 1, Service 3 and Service 4 could be responded accordingly. Hence, the final results have proofed that our mechanism based on check point and TBM is surely effective, i.e., only those trustful access requests defined by TBM could be permitted and responded, while those distrustful requests will be denied automatically. Thus, by means of the mechanism, system can effectively monitor SC's access behaviors in run time.

\begin{figure}[htbp]
\centering
\scalebox{0.33}{\includegraphics{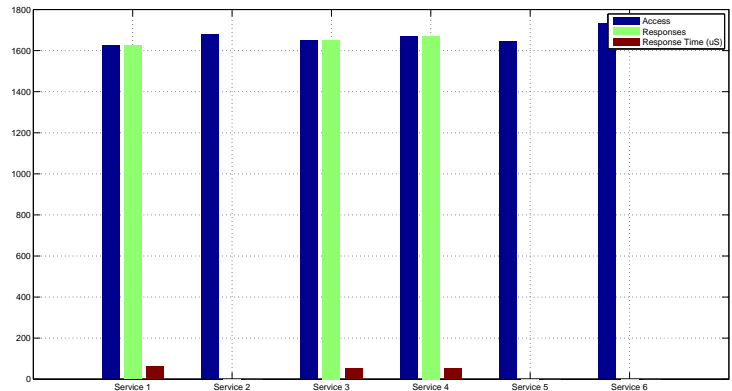}}
\caption{User behavior supervision evaluation under $SRM_{2}$.}
\end{figure}

\subsubsection{Dynamic Access Authorization}

\paragraph{}
\begin{figure}[htbp]
\centering
\scalebox{0.33}{\includegraphics{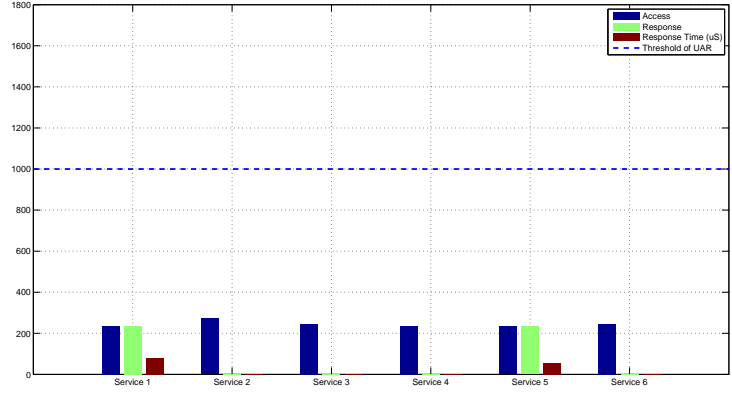}}
\caption{Dynamic access authorization evaluation under $SRM_{1}$ and threshold of UAR is set to 1000 times.}
\end{figure}

Dynamic access authorization evaluation is to validate whether our mechanism can timely terminate a SC's access authorizations when the SC's unauthorized access risk (UAR) or access frequency risk (AFR) really occurs. We first evaluate the mechanism under UAR being triggered. We set the threshold of UAR to 1000 times, i.e., $\theta(UAR)$=1000, which implies that the SC's access authorization will be early terminated in this session when his or her unauthorized access times exceeds 1000 times. Then based on this setting, we repeat the experiments of evaluating user behavior supervision again. When experiments finished, we get the final result and show them in Figure 9 and Figure 10 respectively. We can observe that, compared with the earlier results shown in Figure 7 and Figure 8, the access times in new experiment have dramatically dropped. Taking example of Service 1 in Figure 9, its access times and response times is only 236 times. That is because SC's access authorizations have been early terminated by monitor when unauthorized access times has exceeded 1000 times.

\begin{figure}[htbp]
\centering
\scalebox{0.33}{\includegraphics{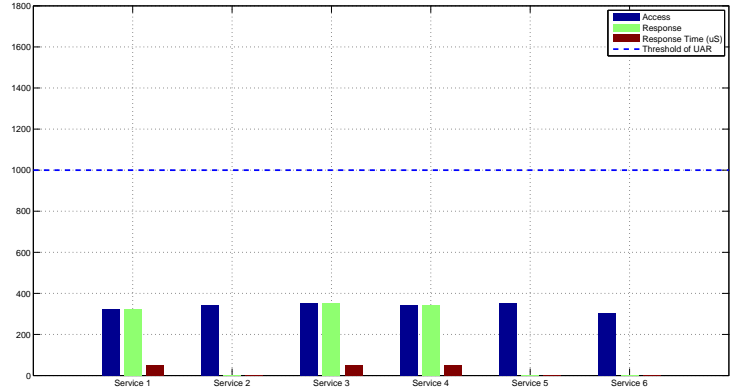}}
\caption{Dynamic access authorization evaluation under $SRM_{2}$ and threshold of UAR is set to 1000 times.}
\end{figure}

\paragraph{}
Next, we evaluate whether our mechanism can make effective decisions when AFR really occurs. We set the threshold of AFR to 350 times per minute, i.e., $\theta(AFR)$=350/min, which implies that when SC's access frequency exceeds 350 times per minute, his or her access authorizations will be early terminated in this session. Here access frequency can be either the access frequency of authorised services or the access frequency of unauthorized services. Specifically, we first randomly produce 20 groups of access frequencies. Then, under the scenario $SRM_{1}$, we access Service 1 with the created random access frequency and record the response time of Service 1 in each group. After experiments finished, we can get the final result and show them in Figure 11. We can observe that, Service 1 can still be responded and we can record the response time in first 6 groups because the access frequency is still under 350 times per minute. But in the 7th group, the frequency is up to 388 times per minute, which has exceeded the threshold, thus AFR is triggered. From the 7th group on, we can't record any response time from Service 1 because SC's access authorization has been early terminated in this session.

\begin{figure}[htbp]
\centering
\scalebox{0.33}{\includegraphics{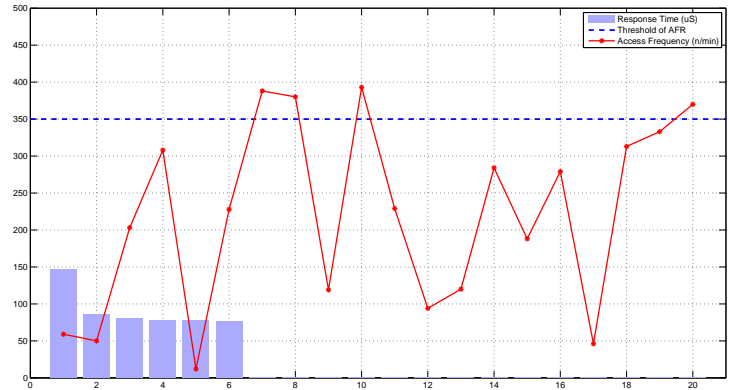}}
\caption{Dynamic access authorization evaluation under $SRM_{1}$ and threshold of AFR is set to 350 times per minute.}
\end{figure}

\paragraph{}
In conclusion, the final experimental results of dynamic access authorization evaluation have proofed that our mechanism is truly effective when UAR or AFR really occurs. Thus, by means of our mechanism, the system can effectively detect a SC's malicious behaviors in run time and make effective response towards these malicious behaviors, i.e., early terminate SC's access authorization in this session.

\subsubsection{Performance Evaluation}

\paragraph{}
The performance evaluation is to validate whether our mechanism can still work well as increasing the scale of $TBM^{C_{i}}$ continuously. The significance of this experiment is that when SOA system model expands, the scale of $TBM^{C_{i}}$ model will be expanded synchronously, thus we want to know whether the system can still work well and its response time can still be acceptable. Specifically, based on the scenario $SRM_{1}$, we carry out 10 groups of experiments, and increase the rule's number of $TBM^{C_{i}}$ from 100 to 1000 with 100 as the basic unit. For each group experiment, we record the response time of accessing Service 1, then we repeat 5 times and calculate an average response time as the final result. After experiments finished, we can get the final result and we plot them in Figure 12 as follows. We can observe that, as increasing the rule's number of $TBM^{C_{i}}$ gradually, the response time of accessing Service 1 is synchronously increased and nearly linear with the number of trustful behavior rules. And all of response time can be acceptable.

\begin{figure}[htbp]
\centering
\scalebox{0.33}{\includegraphics{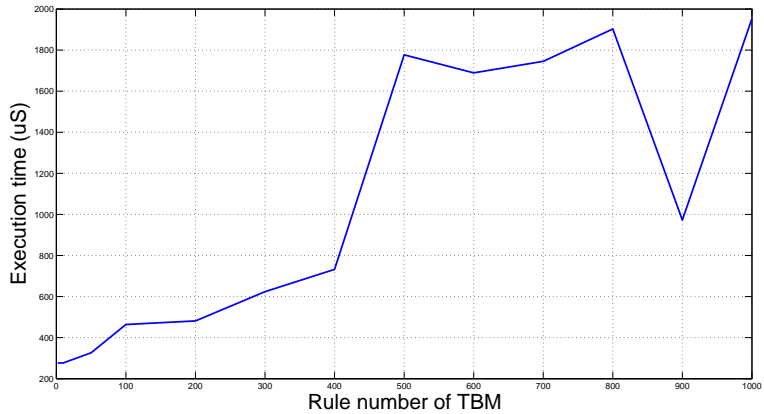}}
\caption{Performance evaluation of accessing Service 1 under $SRM_{1}$.}
\end{figure}

\section{Related Work}

\paragraph{}
After surveying the related works concerning security policy monitoring for SOA systems, we found an important direction in this area is to monitor the business process model and notation (BPMN) \cite{BPMN1} of SOA system. These methods \cite{BPMN2,BPMN3,BPMN4,BPMN5,BPMN5,BPMN6} are based on the runtime monitoring of a service composition to ensure that the service behaves in compliance with a pre-defined security policy. If finding policy violations, the monitor program will send alerts to service provider or system manager. However, these methods didn't consider the security requirements of different users. An ideal policy monitoring might allow different users to be given the opportunity to apply their own security policies enforced through a combination of design-time and run-time checking \cite{BPMN6}. Compared with these methods, our mechanism doesn't need to maintain a BPMN of system, and SRM defined by service provider can be converted into different security policies (i.e TBM) for different service consumers, then monitor program can track the user's behavior using TBM in run time.

\paragraph{}
Branimir et al. \cite{Branimir} proposed a performance monitoring approach for WS-BPEL service compositions. This study depends on business activity monitoring (BAM), which enables continuous, near real-time monitoring of processes based on process performance metrics (PPMs). Business people define PPMs based on business goals, then PPMs are translated to the monitoring models by IT engineers. In case of severe deviations from planned values, alerts are raised and notifications are sent to the responsible people. The similar research works based on BAM are presented in \cite{Ou,B1,B2}. By means of process engine, these works can also find abnormal behaviors, but their monitoring target is of a virtual business process model. Thus, they can only provide near real-time monitoring results.

\paragraph{}
Zuma et al. \cite{zuma} proposed a dynamic monitoring framework for SOA execution environment. The designed framework is based on the monitoring scenario and interceptor mechanism. Monitoring scenario is a group monitoring rules (interceptor chain) defined by service provider and published in the interceptors registry of container. Interceptor socket is an independent service which consists of observer, interceptor chain and data processing engine (CEP). Once a service in SOA system invoked, the socket will be activated at the same time. The collected attributes of the service by observer are send to CEP for analysing. If finding abnormal behaviors, the socket will notify service provider in run time. The similar research work is presented in \cite{Z1}. Although these works can find user's abnormal behaviors in run time, but compared with our mechanism, they lack the associated dynamic response mechanism.

\paragraph{}
Chen et al. \cite{Chen} proposed a framework-based runtime monitoring approach for service-oriented systems. The primary goal of the framework is to support software engineers by creating high-level views of how services interact, i.e., the runtime topology of SOA system. Once the listener of a Web service receives an incoming request, it will forward the message to the proxy, which will parse the request to obtain the information of the invoking service and interface. Then the content of the request is delivered to the target service. After the response sent back, the proxy will stores the addressing information in logging system, by which service provider can find some abnormal behaviors. Although this study can establish a high-level system runtime topology, yet it can not find user's abnormal behaviors automatically in run time. In some sense, it is similar to an offline monitoring mode.

\section{Conclusion}

\paragraph{}
In this paper, we propose a behavior-aware service access control mechanism using security policy monitoring for SOA system.
Firstly, we present a formal system model for modeling SOA system. Then based on the system model, we present our proposed mechanism in detail. We introduce two important formal model, service releasing model and trustful behavior model, in our mechanism. Next, we present a group risk quantitative measurement models for evaluating user's behavior risks in system. Finally, we give the detailed algorithmic description of behavior-aware access control algorithm in our mechanism. In order to evaluate the feasibility of proposed mechanism, we implement a proof-of-concept experimental system. The final experimental results have proofed that our mechanism can effectively monitor SC's access behaviors in run time and truly effective when UAR or AFR really occurs. Thus, by means of our mechanism, the system can effectively detect a SC's malicious or misusing behaviors and make effective dynamic response towards these malicious behaviors in run time, i.e., early terminate SC's access authorization in this session. As increasing the rule's number in TBM continuously, our mechanism can still work well and the response time can be still acceptable.

\paragraph{}
The present work is based on the assumption that one service provider has the ownership of entire SOA system, i.e., all services in SOA system are in one trust domain. But in open cloud computing environment, SOA system always has multiple trust domains. Which implies that different service belongs to different service provider, thus their security policies are also different. Hence, runtime monitoring mechanism for multiple-domains SOA system is a more challenging problem and will be as one important part of our ongoing works.

\section*{Acknowledgments}

\paragraph{}
This paper has been sponsored and supported by National Natural Science Foundation of China (Grant No.61772270), partially supported by National Natural Science Foundation of China (Grant No.61602262).

\section*{References}

\bibliographystyle{elsarticle-num}
\bibliography{m1}



%



\end{document}